\documentclass[fleqn,usenatbib]{mnras}
\usepackage[T1]{fontenc}
\usepackage{ae,aecompl}
\usepackage{booktabs}

\usepackage{graphicx}	
\usepackage{amsmath}	
\usepackage{amssymb}	
\usepackage{natbib}
\bibpunct{(}{)}{;}{a}{}{,} 
\usepackage{xcolor}
\newcommand{\es}[1]{}
\usepackage[caption=false]{subfig}

\newcommand{\highlight}[1]{}

\usepackage{etex}	
\usepackage{acronym}  
\usepackage{float}
\usepackage{makecell}
\usepackage{mathtools}
\usepackage{ulem}
\usepackage{footmisc}

\definecolor{ochre}{rgb}{0.8, 0.47, 0.13}

\title[BPS with stochastic remnant recipes]{Binary population synthesis with probabilistic remnant mass and kick prescriptions}

\author[I.~Mandel et al.]{
	   Ilya Mandel$^{1,2,3}$, 
	   Bernhard M\"{u}ller$^{1,2}$,
	   Jeff Riley$^{1,2}$,
	   Selma E. de Mink$^{4,5}$,\newauthor
	   Alejandro Vigna-G\'{o}mez$^{6}$,
	   Debatri Chattopadhyay$^{7,2}$
\\
$^{1}$ Monash Centre for Astrophysics, School of Physics and Astronomy, Monash University, Clayton, Victoria 3800, Australia\\
$^{2}$ The ARC Center of Excellence for Gravitational Wave Discovery -- OzGrav\\	   
$^{3}$ Birmingham Institute for Gravitational Wave Astronomy and School of Physics and Astronomy,\\ University of Birmingham,  B15 2TT, Birmingham, UK\\
$^{4}$ Center for Astrophysics, Harvard Smithsonian, 60 Garden Street, Cambridge, MA 02138, USA\\
$^{5}$ Anton Pannekoek Institute for Astronomy, University of Amsterdam, Science Park 904, 1098XH, Amsterdam, The Netherlands \\
$^{6}$ DARK, Niels Bohr Institute, University of Copenhagen, Blegdamsvej 17, 2100, Copenhagen, Denmark\\
$^{7}$ Centre for Astrophysics and Supercomputing, Swinburne University of Technology, Hawthorn, VIC 3122, Australia\\
\\
}

\date{Accepted XXX. Received YYY; in original form ZZZ}
\pubyear{2020}

\begin{document}
\label{firstpage}
\pagerange{\pageref{firstpage}--\pageref{lastpage}}
\maketitle

\begin{abstract}
We report on the impact of a probabilistic prescription for compact remnant masses and kicks on massive binary population synthesis.  We find that this prescription populates the putative mass gap between neutron stars and black holes with low-mass black holes.  However, evolutionary effects reduce the number of X-ray binary candidates with low-mass black holes, consistent with the dearth of such systems in the observed sample.  We further find that this prescription is consistent with the formation of heavier binary neutron stars such as GW190425, but over-predicts the masses of Galactic double neutron stars.  The revised natal kicks, particularly increased ultra-stripped supernova kicks, do not directly explain the observed Galactic double neutron star orbital period--eccentricity distribution.  Finally, this prescription allows for the formation of systems similar to the recently discovered extreme mass ratio binary GW190814, but only if we allow for the survival of binaries in which the common envelope is initiated by a donor crossing the Hertzsprung gap, contrary to our standard model.
\end{abstract}


\section{Introduction}\label{sec:intro}

Rapid binary population synthesis models rely on simplified prescriptions to enable computationally efficient modelling of a large set of binaries under a range of assumptions about stellar and binary evolution.  Recipes for predicting the masses and natal kicks of neutron stars or black holes left behind at the end of massive stellar evolution include models by \citet{Hurley:2000,Hobbs:2005,Belczynski:2008,Fryer:2012,Spera:2015,Mueller:2016,BrayEldridge:2016,BrayEldridge:2018,PattonSukhbold:2020}.  Recently, \citet{MandelMueller:2020} proposed a new set of prescriptions, derived from first-principle 3-dimensional supernova simulations and parametrised simplified models. These new recipes are probabilistic (see, e.g., \citealt{Clausen:2015}) in order to account for the stochastic nature of stellar evolution and supernovae.  The \citet{MandelMueller:2020} recipes use the carbon-oxygen core mass of the exploding star to determine the supernova outcome.  Neutron star formation is guaranteed for the lowest carbon-oxygen core masses that allow for supernovae, while sufficiently massive carbon-oxygen cores always yield black hole formation with complete fallback of the core and any remaining helium shell.  Between these regimes, the outcome is probabilistic, including an intermediate possibility of black hole formation with significant supernova mass loss.  In the absence of complete fallback, the remnant mass is a function of the carbon-oxygen core mass, with additional stochastic scatter. These recipes self-consistently couple natal kicks with remnant masses, with stochastic scatter on top of momentum-preserving kicks based on the ratio of the ejected carbon-oxygen core mass (a proxy for the portion of the ejecta that have significant asymmetry) to the remnant mass.  

Here, we investigate the impact of these prescriptions on the evolution of massive stellar binaries using the COMPAS (Compact Object Mergers: Population Astrophysics and Statistics) rapid binary population synthesis suite \citep{Stevenson:2017,VignaGomez:2018}.  We follow the default prescriptions from these papers, with the addition of (pulsational) pair-instability supernovae following the recipes of \citet{Stevenson:2019} and chemically homogeneous evolution following the recipes of \citet{Riley:2020}. We use the \citet{MandelMueller:2020} recipes for compact remnant masses and kicks, with the default parameters from Table 1 in that paper.  We consider solar metallicity ($Z_\odot=0.014$, \citealt{Asplund:2009}) except where $Z=0.1\, Z_\odot$ is specified.   We do not allow donors that are crossing the Hertzsprung gap to survive a common-envelope phase (``Pessimistic'' model of \citealt{Belczynski:2008}).  We assume that any binaries that would experience Roche-lobe overflow immediately after ejecting the common envelope would, in fact, fail to survive the ejection of the envelope and merge during this phase.  When discussing double compact-object binaries, we select only those that will merge within the approximate current age of the Universe, 14 Gyr, through gravitational-wave emission.  Primary masses are sampled from a Kroupa initial mass function with secondary masses following a flat distribution in the mass ratio, while separations of initially circular binaries follow a flat-in-log distribution \citep{Opik:1924,Kroupa:2001,Sana:2012}.  A total of 1 million binaries are evolved with zero-age main-sequence primary masses between 5 and 150 M$_\odot$ and initial separations between 0.01 and 1000 au, representing $\sim 90$ million solar masses of total star formation, at each of $Z_\odot$ and $0.1\, Z_\odot$.

\section{Mass gap and Black-hole X-ray binaries}\label{sec:XRB}

Observations of Galactic neutron stars suggest a likely maximum mass of $\approx 2.1$ M$_\odot$ \citep{Antoniadis:2016,Alsing:2018,Farr:2020}.  This is consistent with the maximum stable neutron star mass inferred from observations of the double neutron star merger GW170817 \citep{MargalitMetzger:2017} and the maximum theoretically possible mass without relying on exotic matter \citep[e.g.,][]{Lattimer:2012}.  On the other hand, masses of black holes in black hole X-ray binaries are consistent with a minimal mass of $\approx$ 4.5--5 solar masses \citep{Ozel:2010,Farr:2010}.  This led to the conjecture of a mass gap between neutron stars and black holes, and the emergence of remnant mass models that enforced such a mass gap, e.g., the ``Rapid'' model of \citet{Fryer:2012}.

However, more recent evidence indicates that the putative mass gap is at least partly populated.  Microlensing observations hint at a population of dark remnants in the mass gap \citep{WyrzykowskiMandel:2019}.  Several recent observation point to non-interacting binaries with dark companions that are likely to fall in the mass gap \citep[e.g.,][]{Thompson:2019}.   The best-measured compact object in the mass gap is the lighter companion in the recent gravitational-wave observation GW190814, with a mass of $2.59^{+0.08}_{-0.09}\ \mathrm{M}_\odot$ \citep{GW190814}.  On the other hand, supernova models suggest that the remnant mass range of the mass gap could be reasonably well populated \citep[e.g.,][]{Ertl:2020,Chan:2020}.

Therefore, to satisfy the available evidence, a remnant mass model must both populate the mass gap and be consistent with the dearth of observations of black hole X-ray binaries with low-mass black holes.  \citet{MandelMueller:2020} demonstrated that, following their remnant mass prescription, a population of single stars drawn from the initial mass function will populate the mass gap with compact remnants, which would, for example, explain the microlensing observations (the `Delayed' model of \citealt{Fryer:2012} also populates the mass gap).  We will return to merging compact-object binaries and gravitational-wave sources in Section \ref{sec:GW}.  Here, we consider the impact of the \citet{MandelMueller:2020} remnant mass and kick model on X-ray binary predictions.

We identify any binaries in which one star collapses into a black hole while the companion is non-degenerate as potential X-ray binaries.  In our binary population synthesis models, the most typical companion at the time the black hole is formed (in more than 80\% of such systems) is a main-sequence star of above 3 solar masses (i.e., O or B star).  

Approximately a fifth of all black holes formed in binaries with a non-degenerate companion are low-mass black holes, which we conservatively define as having a mass between 2 and 4 solar masses.  At first glance, this would suggest that around a fifth of black-hole X-ray binaries should host low-mass (or mass-gap) black holes.  

\begin{figure}
\centering
	\includegraphics[width=0.45\textwidth]{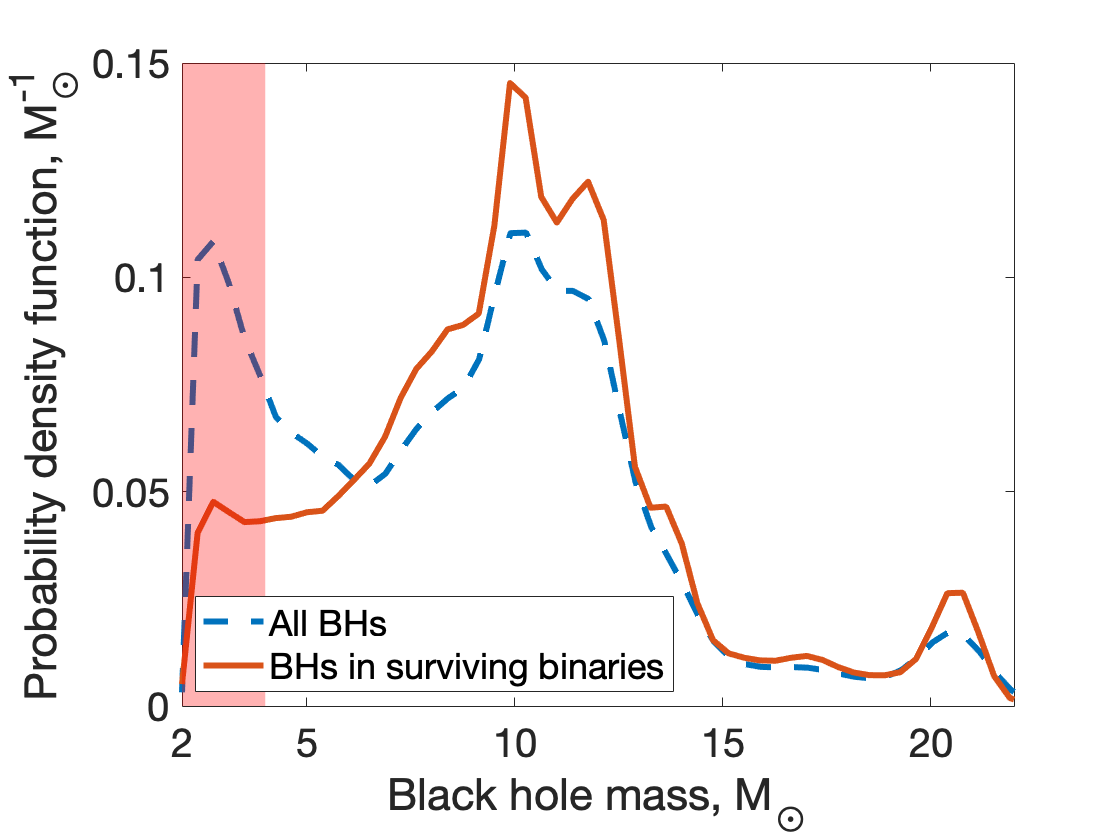}
	\caption{Mass distribution of black holes at formation in binaries where the companion is a non-degenerate star.  The two distributions show all binaries, including those disrupted by the supernova  (dashed) and those binaries that survive the black hole formation and associated kick (solid).  The shaded region indicates low-mass (mass-gap) black holes.  Simulations at $\mathrm{Z}_\odot$ are shown.}
	\label{fig:BHmass}
\end{figure}

However this does not account for the disruption of binaries due to mass loss during a supernova or natal kicks.  The birth of low-mass black holes is accompanied by the highest amount of mass loss and, in the \citet{MandelMueller:2020} models (see also \citealt{Fryer:2012,BrayEldridge:2016,BrayEldridge:2018,GiacobboMapelli:2020}), the largest natal kicks.  Consequently, binaries with low-mass black holes are the most likely to be disrupted by supernovae, as shown in Figure \ref{fig:BHmass}.  In fact, while the vast majority ($\approx 97\%$) of binaries that form $\geq 5\ \mathrm{M}_\odot$ black holes survive the first supernova, only just over a third of binaries that form low-mass black do so.  Therefore, only around 8\% of surviving black-hole X-ray binary progenitor candidate binaries host low-mass black holes.  

Thus, only one or two of the 20 black-hole X-ray binaries with well-measured masses analysed by \citet{Farr:2010} should be expected to host a black hole in the putative mass gap.  At least one of those binaries may, in fact, be consistent with a low-mass black hole once possible measurement biases are accounted for \citep{Kreidberg:2012}.  Even if none are, it is hardly surprising: the probability of this is $(1-0.08)^{20} \approx 19\%$.

We only considered potential X-ray binaries and intentionally avoided the discussion of observability, since that would require further assumptions to be made on the duration and luminosity of X-ray emission.  However, it is not unreasonable to expect that observational selection effects would further diminish the contribution of X-ray binaries with low-mass black holes to the observed sample.  For example, not surprisingly, a higher fraction of potential systems with low-mass black holes have companions more massive than the black hole after the supernova.  These would then be detectable as high-mass wind-fed X-ray binaries, which are shorter-lived than stably mass transferring low-mass X-ray binaries. At the very least, the expected maximum Eddington luminosity is lower for low-mass black holes.

\section{Double neutron stars}\label{sec:DNS}

Multiple binary population synthesis studies explored the observed properties of Galactic double neutron star systems, including \citet{1998AA...332..173P,Oslowski:2011,Andrews:2015,Tauris:2017,ShaoLi:2018,VignaGomez:2018,Kruckow:2018}.  However, while most of these studies are able to reproduce individual binaries, they struggle to accurately reproduce the distribution of the properties of neutron star binaries in the population.   In particular, models struggle to match the observed distribution of double neutron star masses \citep[e.g.,][]{Tauris:2017,VignaGomez:2018,Kruckow:2020} and eccentricities \citep[e.g.,][]{Ihm:2006,Chruslinska:2017,AndrewsMandel:2019,Debatri:2020}.  Unfortunately, the stochastic remnant mass and kick recipe \citep{MandelMueller:2020} is not able to resolve the discrepancy between predictions and observations without further modifications to the model assumptions.

The first-born neutron star is formed in an electron-capture supernova in more than half of merging double neutron stars in our simulations, giving it a birth mass of $1.26\ \mathrm{M}_\odot$.  Subsequent Eddington-limited accretion raises its mass by no more than $0.001\ \mathrm{M}_\odot$, producing a sharp peak in the model distribution which is not seen in observations.  This suggests that, contrary to our models, there is either a broader range of electron-capture supernova birth masses, or significant super-Eddington accretion, perhaps during the common-envelope phase \citep[e.g.,][]{MacLeodRamirezRuiz:2015} or during mass transfer from a post-He-main-sequence secondary onto the neutron star primary (case BB mass transfer, \citealt{DewiPols:2003}). 

\begin{figure}
\centering
	\includegraphics[width=0.45\textwidth]{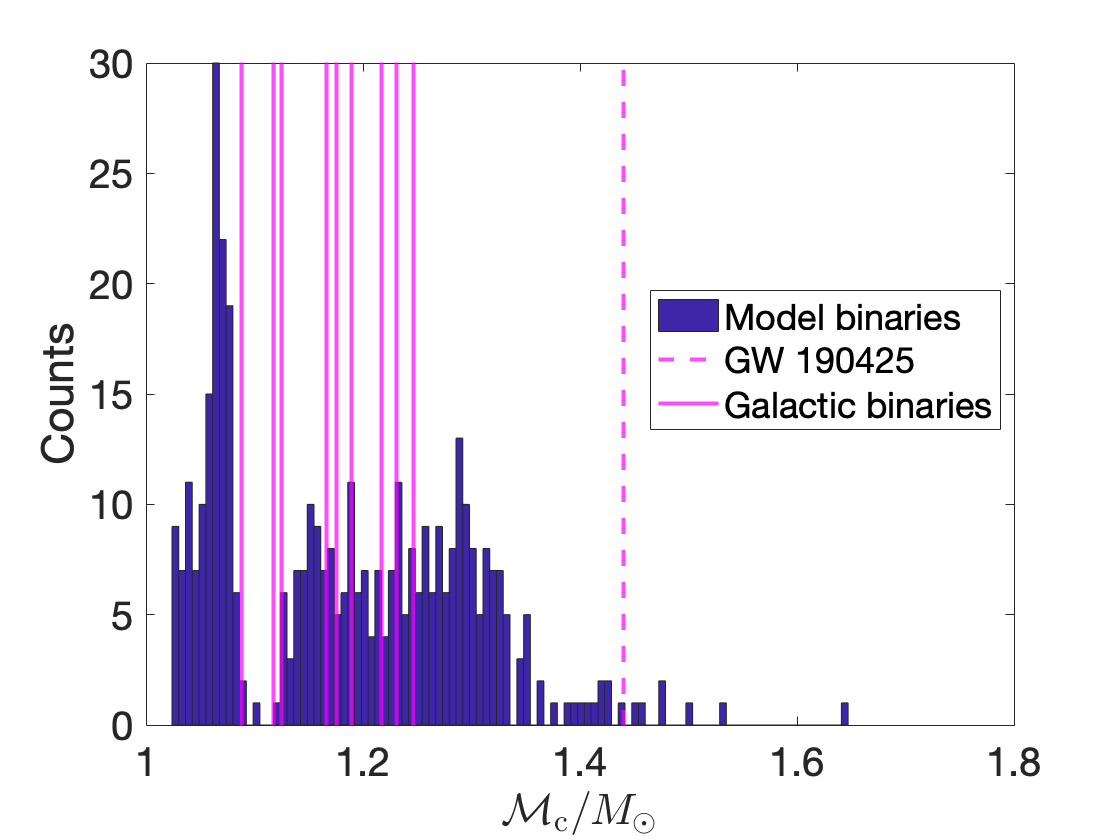}
	\caption{Chirp mass distribution of double neutron stars formed at solar metallicity and merging within 14 Gyr.  The magenta lines indicates the chirp masses of Galactic double neutron stars (solid) and GW190425 (dashed).}
	\label{fig:DNSmass}
\end{figure} 

Moreover, we find that the chirp mass, defined as ${\mathcal{M}}_{\mathrm{c}} = m_1^{3/5} m_2^{3/5} (m_1+m_2)^{-1/5}$, where $m_{1,2}$ are the individual neutron star masses, is generally over-predicted by the model relative to the observations.  Figure \ref{fig:DNSmass} shows the predicted chirp mass distribution of merging neutron stars, which we focus on because the first-born pulsars in merging binaries are at least mildly recycled, while wider binaries are more likely to suffer from challenging selection effects \citep{Debatri:2020}, and to allow for comparison with gravitational-wave sources.  A third of all simulated systems have a chirp mass higher than that of PSR J1913 + 1102, the Galactic system with the highest chirp mass among the nine systems with precisely measured masses (see, e.g., the compilation of \citealt{Farrow:2019}).  The probability of this happening by chance (ignoring any selection effects, cf.~\citealt{Debatri:2020}) is only 3\%.  One possible correction is that binary evolution changes the compactness of the core, leading to a reduced remnant mass relative to single stars with the same carbon-oxygen core mass for which the \citet{MandelMueller:2020} model was calibrated.  This could reduce the predicted masses by a few hundredths of a solar mass, bringing them into closer agreement with observations.  

On the other hand, we find that massive binaries such as GW190425, a binary neutron star merger with a chirp mass of $1.44 \pm 0.02\ \mathrm{M}_\odot$, could be generically produced through binary population synthesis.  Among all merging binary neutron stars in our models, 3\% have a chirp mass exceeding 1.42 $\mathrm{M}_\odot$ at solar metallicity, rising to 8\% at $0.1\ \mathrm{Z}_\odot$.  This increases to 4\% (12\%) at $\mathrm{Z}_\odot$ ($0.1\ \mathrm{Z}_\odot$) after accounting for gravitational-wave selection effects, which scale as ${\mathcal{M}}_{\mathrm{c}}^{2.5}$ since the detection range scales as ${\mathcal{M}}_{\mathrm{c}}^{5/6}$.  There is then a $\approx 8\%$ (21\%) probability that one of two reported gravitational-wave binary neutron stars is as massive as GW190425 (the other, GW170817 \citep{GW170817}, had a mass similar to Galactic double neutron stars).  We thus do not find the need to resort to alternative models \citep[e.g.,][]{RomeroShaw:2020,Safarzadeh:2020} to explain GW190425; see Vigna-G\'{o}mez, Ramirez-Ruiz and Mandel (in prep.), for more details.

\citet{AndrewsMandel:2019} argued that the default model of binary evolution will naturally struggle to reproduce the observed bi-modal distribution of eccentricities of short-period Galactic double neutron stars.  The peak around eccentricity $0.1$ is consistent with ultra-stripping during case BB mass transfer.  This leaves behind a low-mass star that easily explodes in an ultra-stripped supernova with limited mass loss and without significant asymmetry, giving rise to reduced kicks \citep{Tauris:2015}.  However, it is challenging to explain the cluster of three Galactic binary neutron stars, including the Hulse-Taylor binary B1913+16, with eccentricities around 0.6, since such systems should have also experienced ultra-stripped supernovae, or the absence of observed binaries with intermediate eccentricities.

The stochastic kick model of \citet{MandelMueller:2020} does not treat ultra-stripped supernovae separately from other supernova types (see also \citealt{BrayEldridge:2016,GiacobboMapelli:2020}).  Instead, the kick is proportional to the mass of the carbon-oxygen core ejected during the supernova, which characterises the asymmetric ejection of linear momentum, and inversely proportional to the remnant mass.  This reduces the ultra-stripped supernova kick relative to typical core collapse supernovae, and puts it in line with low-mass iron core collapse \citep{Podsiadlowski:2004,Mueller:2018,Stockinger:2020}.  It nevertheless allows for mean natal kicks of $\sim 200$ km s$^{-1}$ for ultra-stripped neutron stars, significantly greater than the 30 km s$^{-1}$ ultra-stripped supernova kicks assumed in previous studies such as \citet{VignaGomez:2018}.  For comparison, mean natal kicks for electron capture supernovae are only $\sim 40$ km s$^{-1}$ in the solar-metallicity models considered here, while those associated with neutron star birth through regular core collapse supernovae are $\sim 300$ km s$^{-1}$.  The latter value is smaller than the \citet{Hobbs:2005} estimate of observed pulsar speeds (mean of $\sim 400$ km s$^{-1}$), but see \citet{MandelMueller:2020} for a discussion of possible evolutionary and observational selection effects.

\begin{figure} 
\centering
	\includegraphics[width=0.45\textwidth]{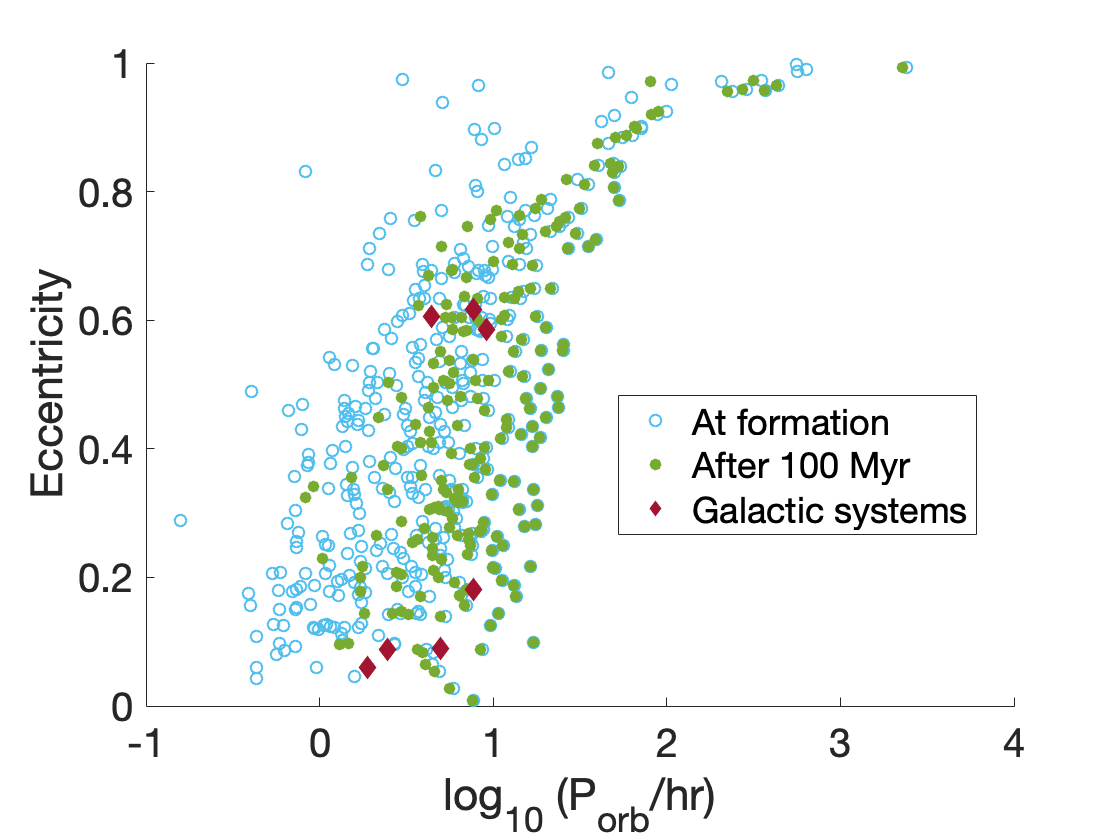}
	\caption{Period-eccentricity distribution of double neutron stars merging within 14 Gyr at the moment of formation (open blue circles) and after the passage of 100 Myr (green solid circles -- note that merged systems no longer appear on the plot).  Merging field Galactic double neutron stars are shown with burgundy diamonds.}
	\label{fig:pe}
\end{figure} 

We may thus expect binary population synthesis based on the  \citet{MandelMueller:2020} models to predict a broader range of eccentricities for short-period neutron stars, and that is indeed what we see in Figure \ref{fig:pe}.  We find a broad, relatively flat distribution of eccentricities of merging double neutron stars, both at birth and after 100 Myr of orbital decay through gravitational-wave emission, a typical spin-down age for mildly recycled pulsars in the observed sample.  However, the modelled distribution does not show two clusters at eccentricities near $0.1$ and $0.6$ that appear in the observed population.  Moreover, the low-eccentricity population is depleted relative to \citet{VignaGomez:2018} models by the increased ultra-stripped supernova kicks.  We thus conclude that there is missing physics in the model if, in fact, all observed field double neutron stars are formed through the isolated binary evolution channel.  One possibility could be the impact of incomplete stripping of the hydrogen envelope during earlier evolutionary phases and enhanced envelope re-expansion after the end of core helium burning in partially stripped stars \citep{Gotberg:2017,Laplace:2020}.

\section{Merging binaries}\label{sec:GW}

\begin{table*}
\centering
\begin{tabular}{|l|l|l|}
\hline
Binaries  & \multicolumn{2}{c|}{Yield, $\times 10^{-6}\ \mathrm{M}_\odot^{-1}$}\\
& at $\mathrm{Z}_\odot$\ \ \ \ & at $0.1\ \mathrm{Z}_\odot$\\
\hline
Binary neutron stars & 4.6 & 2.0\\
Black hole -- neutron star binaries (all) & 16 & 9.0\\
--- with $2\ \mathrm{M}_\odot < M_\mathrm{BH} \leq 4\ \mathrm{M}_\odot$ & 4.7 & 1.7\\
Binary black holes (all) & 11 & 67\\
--- with $2\ \mathrm{M}_\odot < \min(M_\mathrm{BH}) \leq 4\ \mathrm{M}_\odot$ & 6.2 & 4.2\\
\hline
\end{tabular}
\caption{Yields of compact-object binaries that will merge in 14 Gyr per million solar masses of star formation.}\label{table:rates}
\end{table*}

Finally, we consider predictions for merging compact-object binaries.  Table \ref{table:rates} summarises the predicted yields per million solar masses of star formation at two metallicities: solar and a tenths of solar.  We separately highlight the formation of merging (within 14 Gyr) neutron star -- black hole binaries and binary black holes that include a low-mass (mass-gap) black hole as one of the companions.  

The overall yields are broadly consistent with previous COMPAS studies, such as \citet{Neijssel:2019}.  We find a reduced yield of binary neutron stars relative to that study, but this is largely a consequence of a correction to the treatment of the common envelope phase in COMPAS  \citep{VignaGomez:2020} rather than a change in the remnant mass and kick physics.  Unlike that study, we find that binaries involving neutron stars have reduced yields at lower metallicity (but see \citealt{Laplace:2020}). Meanwhile, metallicity has a very large impact on the progenitors of binary black holes which suffer from strong winds at high metallicity, reducing the binary mass and widening the binary so much that it can no longer merge.  This is compounded by the impact of metallicity on the radial expansion of massive stars, with a reduced window for radial growth during core helium burning that is necessary to tighten the binary through a survivable common envelope in the COMPAS models \citep[e.g.,][]{Klencki:2020}, leading to an order-of-magnitude reduction in the yield of merging binary black holes at $\mathrm{Z}_\odot$ relative to $0.1 \mathrm{Z}_\odot$.  We do find increased binary black hole yields at both $\mathrm{Z}_\odot$ and $0.1 \mathrm{Z}_\odot$ with the \citet{MandelMueller:2020} remnant mass and kick prescription relative to the \citet{Fryer:2012} `Delayed' prescription (cf.~Figure 7 of \citealt{Riley:2020}) when using the same mass threshold of $2.0\, M_\odot$ between neutron stars and black holes.

The recently published gravitational-wave discovery GW190814 is a signal from the merger of a $\approx 2.6\ \mathrm{M}_\odot$  compact object, which we assume to be a black hole, with a $\approx 23\ \mathrm{M}_\odot$ black hole \citep{GW190814}.  This source is challenging to form in sufficient quantities  through isolated binary evolution \citep{Zevin:2020}, and alternative models have been proposed ranging from formation in young stellar clusters \citep{Rastello:2020} or quadruple dynamics \citep{Fragione:2020} to less standard models such as strong lensing \citep{Broadhurst:2020} or growth by accretion from a circumbinary ring \citep{Safarzadeh:2020fallback}.  

We also struggle to reproduce this source, despite naturally allowing for black holes with the right masses.  While binary black holes with unequal mass ratios are common in our simulations, with a median mass ratio of 0.6 and mass ratios below $0.3$ in 15\% of merging binaries at $\mathrm{Z}_\odot$, only 2\% of merging binary black holes at $\mathrm{Z}_\odot$ have mass ratios below $0.2$, and none are as extreme as GW190814.  While the overall yield of merging binary black holes increases by a factor of six at $0.1 \mathrm{Z}_\odot$ relative to $\mathrm{Z}_\odot$, the yield of binary black holes including a low-mass companion is, in fact, reduced with our default settings.

We can reproduce systems like GW190814 at $0.1 \mathrm{Z}_\odot$ if we allow Hertzsprung-gap donors to survive common-envelope evolution.  In that case, we find a number of systems similar to Channel B of \citet{Zevin:2020}: $\approx 3\times10^{-8}$ merging binary black holes with one companion above 20 M$_\odot$ and the other between 2 and 3 M$_\odot$ per solar mass of star formation.  For example, a binary with initial masses of approximately 40 and 30 $\mathrm{M}_\odot$ at a separation of $\lesssim 30 R_\odot$ undergoes conservative mass transfer while both stars are still on the main sequence, with mass ratio reversal gradually leading into continuing stable mass transfer from the Hertzsprung gap primary, widening the binary.  In the model of \citet{Hurley:2000}, the primary is left with a sufficiently low-mass core to ultimately leave behind a $2.6\ \mathrm{M}_\odot$ black hole, although population synthesis models may under-predict the masses of stripped donor stars that engaged in mass transfer on the main sequence.
Meanwhile, the secondary evolves off the main sequence and, as a Hertzsprung gap donor, initiates dynamically unstable mass transfer onto the primary.  After the binary is hardened by ejecting the common envelope, the secondary collapses into a 23 $\mathrm{M}_\odot$ black hole.  While the survival of common-envelope evolution initiated by a Hertzsprung gap donor is inconsistent with our default model, this mass transfer could, in fact, be dynamically stable based on updated models of the radial response of Hertzsprung-gap donors to mass loss \citep{Ge:2020}; see Neijssel et al.~(in prep.) for further discussion.   Since the secondary is significantly more massive than the primary, and Eddington-limited mass transfer onto a black hole is expected to be almost completely non-conservative, such stable mass transfer could harden the binary sufficiently to allow gravitational-wave emission to drive it to merger within 14 Gyr.  

\section{Conclusion}\label{sec:conc}

We investigated the impact of a probabilistic recipe for neutron star and black hole masses and kicks \citep{MandelMueller:2020} on massive binary evolution.  We found that, as expected, this recipe produces binaries with low-mass black holes in the putative mass gap, consistent with the latest observations, including GW190814.  At the same time, the natural coupling of the remnant kicks and masses helps explain the dearth of low-mass black holes in observed black-hole X-ray binaries.  The recipe is consistent with the formation of more massive binary neutron stars such as GW190425.  However, we continue to struggle to explain the mass distribution and period--eccentricity distribution of Galactic double neutron stars, indicating a shortcoming in binary evolution models or the presence of multiple formation channels for these systems.

\section*{Acknowledgements}
Simulations in this paper made use of the COMPAS rapid population synthesis code (version 2.15.11), which is freely available at \url{http://github.com/TeamCOMPAS/COMPAS}.
The authors thank colleagues in Team COMPAS for helpful discussions, and Tim Riley for assistance in running COMPAS simulations.
IM is a recipient of the Australian Research Council Future Fellowship FT190100574.  
SdM acknowledges funding by the European Union's Horizon 2020 research and innovation program (ERC Grant No.~715063), and by the Netherlands Organization for Scientific Research (Vidi Grant No.~639.042.728).

\section*{Data availability}
The data underlying this article will be available via \url{https://zenodo.org/communities/compas/}.


\bibliographystyle{mnras} 
\bibliography{Mandel}
    
\label{lastpage}
\end{document}